\documentclass{PoS}

\usepackage{bm}
\usepackage{mathptmx}
\usepackage{amsmath}
\usepackage{bm}
\usepackage{textcomp}

\newcommand{\dd}{\mbox{d}} 

\DeclareMathAlphabet\mathbfcal{OMS}{cmsy}{b}{n}

\title{Detector efficiency and exposure of Tunka-Rex for cosmic-ray air showers}

\ShortTitle{Detector efficiency and exposure of Tunka-Rex for cosmic-ray air showers}

\author{
\speaker{O.~Fedorov}$^{1}$, P.A.~Bezyazeekov$^{1}$, N.M.~Budnev$^{1}$, D.~Chernykh$^{1}$, O.A.~Gress$^{1}$, A.~Haungs$^{2}$, R.~Hiller$^{2}$\thanks{now at the University of Z\"urich}, T.~Huege$^{2}$, Y.~Kazarina$^{1}$, M.~Kleifges$^{3}$, E.E.~Korosteleva$^{4}$, D.~Kostunin$^{2}$, O.~Kr\"omer$^{3}$, L.A.~Kuzmichev$^{4}$, V.~Lenok$^2$, N.~Lubsandorzhiev$^{4}$, T.~Marshalkina$^{1}$, R.R.~Mirgazov$^{1}$, R.~Monkhoev$^{1}$, E.~Osipova$^{4}$, A.~Pakhorukov$^{1}$, L.~Pankov$^{1}$, V.V.~Prosin$^{4}$, F.G.~Schr\"oder$^{2}$, A.~Zagorodnikov$^{1}$
-- 
Tunka-Rex Collaboration \\
\llap{$^1$} Applied Physics Institute of Irkutsk State University, Irkutsk, Russia\\
\llap{$^2$} Institut f\"ur Kernphysik, Karlsruhe Institute of Technology (KIT), Karlsruhe, Germany\\
\llap{$^3$} Institut f\"ur Prozessdatenverarbeitung und Elektronik, Karlsruhe Institute of Technology (KIT), Karlsruhe, Germany\\
\llap{$^4$} Skobeltsyn Institute of Nuclear Physics MSU, Moscow, Russia\\
E-mail: \email{offedoroff@yandex.ru}       
}

\abstract{
Tunka-Rex (Tunka Radio Extension) is an antenna array for cosmic-ray detection located in Siberia. 
Previous studies of cosmic rays with Tunka-Rex have shown high precision in determining the energy of the primary particle and the possibility to reconstruct the depth of the shower maximum. 
The next step is the reconstruction of the mass composition and the energy spectrum of cosmic rays. 
One of the main problems appearing within this task is to estimate the detection efficiency of the instrument, and the exposure of the observations. 
The detection efficiency depends on properties of the primary cosmic rays, such as energy and arrival direction, as well as on many parameters of the instrument: 
density of the array, efficiency of the receiving antennas, signal-detection threshold, data-acquisition acceptance, and trigger properties. 
More than that, the configuration of detector changes with time. 
During the measurements some parts of the detector can provide corrupted data or sometimes do not operate. 
All these features should be taken into account for an estimation of the detection efficiency. 
For each energy and arrival direction we estimate the detection probability and effective area of the instrument.
To estimate the detection probability of a shower we use a simple Monte Carlo model, which predicts the size of the footprint of the radio emission as function of the primary energy and arrival direction (taking into account the geometry of Earth's magnetic field). 
Combining these approaches we calculate the event statistics and exposure for each run. 
This is the first accurate study of the exposure for irregular large-scale radio arrays taking into account most important features of detection, which will be used for the measurement of primary cosmic-ray spectra with Tunka-Rex.
}

\FullConference{35th International Cosmic Ray Conference -- ICRC2017\\
		10-20 July, 2017\\
		Bexco, Busan, Korea}

\begin{document}


\section{Introduction}
Tunka-Rex (Tunka Radio Extension)~\cite{TunkaRex_NIM_2015} is a cosmic-ray radio detector, part of the Tunka Advanced Instrument for cosmic ray physics and Gamma Astronomy (TAIGA)~\cite{Budnev:2017fyg}, 
located in Siberia close to Lake Baikal.
The detector antenna array consists of 63 antenna stations with two perpendicularly aligned active antennas of the SALLA type distributed over an area of 3~km$^2$ with a 1~km$^2$ dense core containing 57 stations in the center of the array.
For the time being Tunka-Rex is triggered by Tunka-133 air-Cherenkov detector~\cite{Prosin:2015voa} during winter moonless nights and by Tunka-Grande scintillators the rest of the time~\cite{Budnev:2015cha}.

Tunka-Rex has proven a high resolution for the reconstruction of the energy and the shower maximum~\cite{Bezyazeekov:2015ica}. The next natural step is measuring cosmic-ray spectra with radio detector.
In the last two years the radio detector was significantly upgraded~\cite{Kostunin:2017bzd}: the density of antennas was increased by three times by installation of new antennas, and the duty-cycle reached almost around-the-clock by connecting antennas to the recently installed particle detector Tunka-Grande.

Since one of the objectives of the Tunka-Rex instrument is the observation of the primary cosmic-ray energy spectrum, the issues related to the exposure calculation for the real detector become of great importance.
The problem is divided in two parts: determining the instrument operation time and obtaining the correct aperture of the detector.
To solve that problem use is made of an approach of accurate analysis of the instrument operation during the time.

In case of the Tunka-Rex detector this problem is more complex than for typical cosmic-ray detectors.
Firstly, the intensity of the radio emission depends on the arrival direction, which means, that sensitivity of the detector changes not only with zenith (as for most of cosmic-ray setups), but also with the azimuth angle of incoming cosmic ray because of the geometric nature of the radio emission.
Secondly, the Tunka-Rex array has an irregular structure (antenna stations grouped by clusters with various distances between antenna stations inside the cluster) changing with time due to array upgrades.
Finally, it is also important to take into account any malfunctions and hardware downtimes.
\section{Detector aperture} 
\label{sect3_1}
The exposure $\mathcal{E}$ of the detector is calculated as~\cite{Abreu2011368}:

\begin{equation}
\mathcal{E}(E) = \int_{t}\int_{\Omega}\int_{S}\varepsilon(E, t, \theta,\phi, x,y) \cos\theta\, \dd S\, \dd \Omega\, \dd t = \int_{t}\mathcal{A}(E,t) \dd t
\end{equation}

where $\varepsilon$ is the detection efficiency for events with primary energy $E$, arrival direction $(\theta, \phi)$ and core $(x,y)$,
$t$ are the total time of measurements, 
$\Omega$ and $\dd \Omega = \sin\theta \dd\theta \dd\phi$ is the total and differential solid angles, respectively,
and $S$ is effective surface area of the detector.
The instantaneous aperture of detector $\mathcal{A}(E,t)$ depends on the detector configuration at the time $t$.

The main difference to particle-detector arrays in the definition of detection efficiency\\ ${\varepsilon(E, t, \theta,\phi, x,y)}$ is that the dependence on the azimuth angle cannot be neglected (as for optical and particle detectors),
because the amplitude of the geomagnetic radio emission is proportional to $\sin\alpha$, where ${\alpha = \alpha(\theta,\phi,\theta_B,\phi_B)}$ is the geomagnetic angle, between the shower axis $(\theta,\phi)$ and Earth's magnetic field $(\theta_B,\phi_B)$.
Thus, hereafter the detection efficiency is considered as a function of $\theta$ and $\alpha$, i.e. ${\varepsilon = \varepsilon(E, t, \theta, \alpha, x,y)}$.

\begin{figure}[t]
\centering
\includegraphics[width=0.33\linewidth]{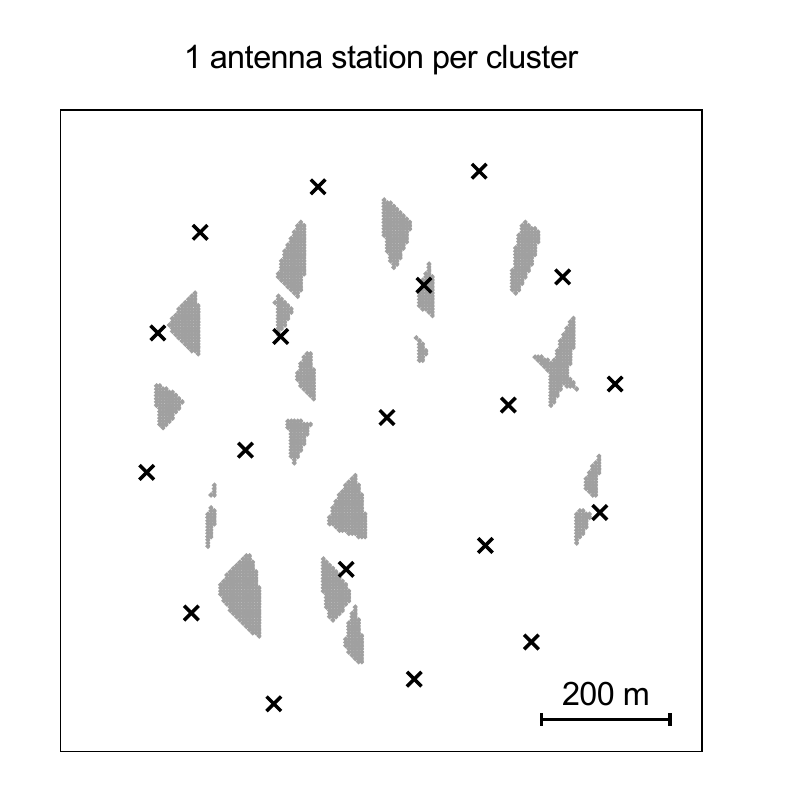}~
\includegraphics[width=0.33\linewidth]{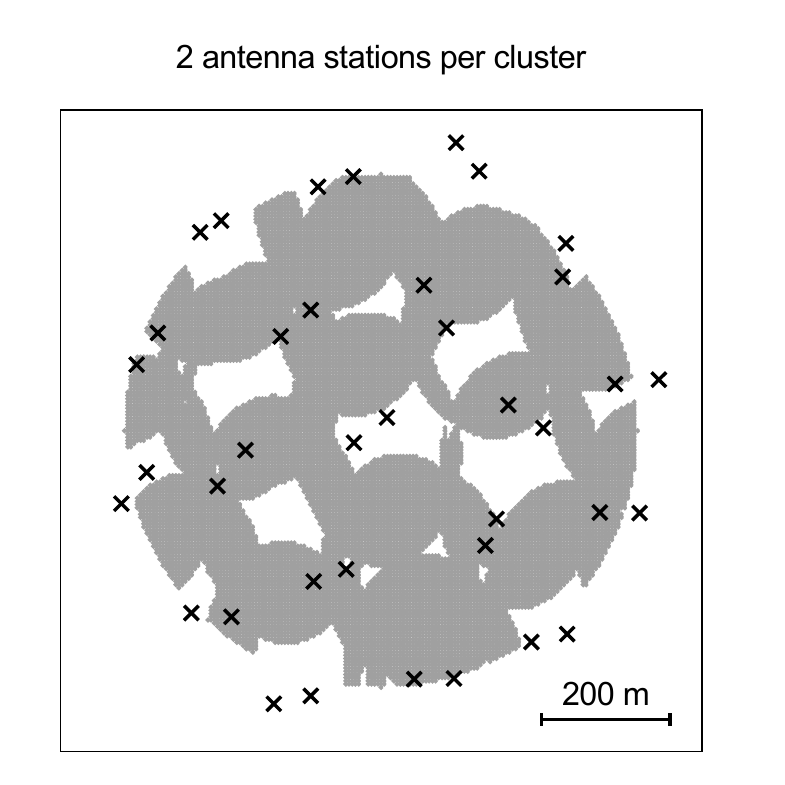}~
\includegraphics[width=0.33\linewidth]{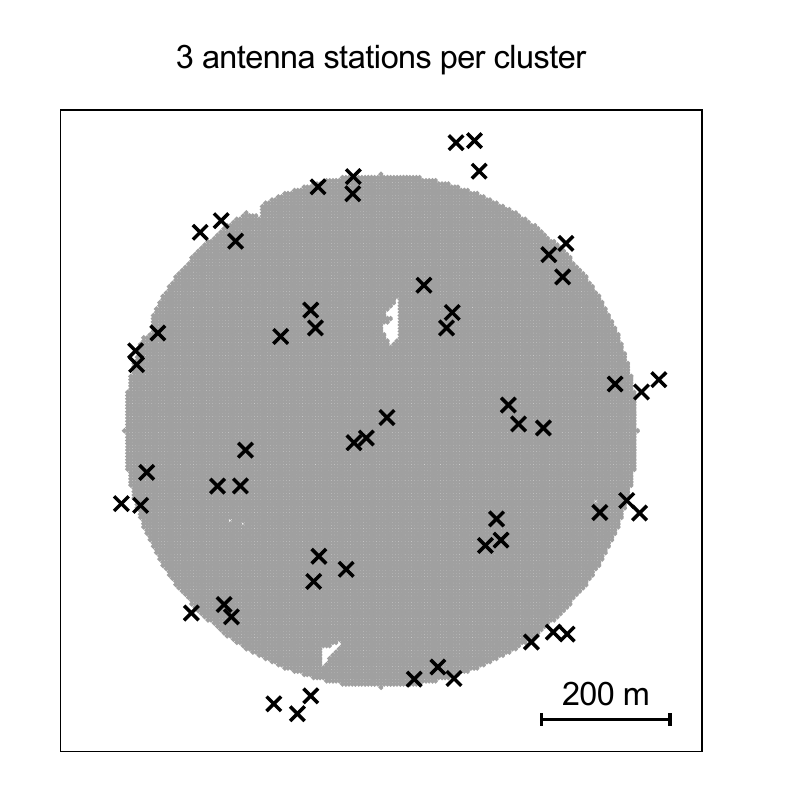}
\caption{
The distribution of the cores of simulated events with ${E = 10^{17.3}}$~eV, ${\theta = 45^\circ}$ and ${\alpha = 27^\circ}$ within a circle of 400~m detected by the Tunka-Rex array with 1, 2 or 3 antenna station per cluster (left to right).
The aperture is not radially symmetric because of the arrival direction, and increases with the number of antennas per cluster.
}
\label{fig:cores_distro}
\end{figure}

To calculate the Tunka-Rex efficiency, a toy Monte Carlo approach has been used (for description see Ref.~\cite{HillerThesis2016}).
Based on a phenomenological model fitting Tunka-Rex measurements, the air-shower footprint above threshold is calculated as function of $E$, $\theta$, $\phi$.
Then these footprints are uniformly distributed on the surface of detector, and only those, which hit three and more antenna stations are considered as detected, i.e.
\begin{equation}
\varepsilon = \varepsilon(E, t, \theta, \alpha, x,y) = \frac{N_\mathrm{detected}}{N_\mathrm{total}}
\end{equation}
The simulated distributions are given in Fig.~\ref{fig:cores_distro}.
One can see, that there are blank spaces in aperture indicating places of the detector with less sensitivity.
These places can be treated separately, 
however, noise and shower-to-shower fluctuations influence to the sensitivity of the entire detector.
By this, the shower core coordinates inside the effective area are not considered and efficiency is averaged over the surface of the antenna array.

\begin{figure}[t]
\centering
\includegraphics[width=1.0\linewidth]{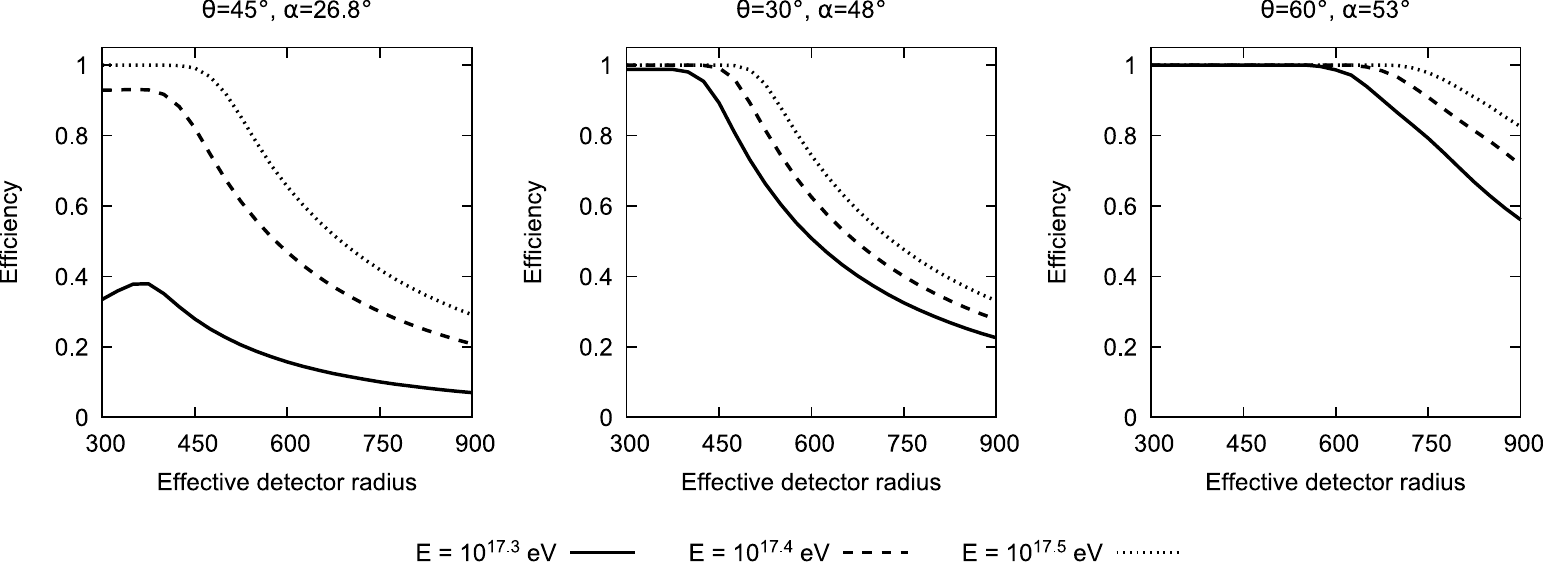}
\caption{The simulated dependence of the efficiency on the effective detector radius $R$ for the different energies and arrival direction of cosmic rays.
One can see the fall of the efficiency after a certain radius, and the effective radius is defined as the edge where this fall starts.}
\label{fig:eff-radius}
\end{figure}

In the case of Tunka-Rex detector, the efficiency can be factorized as follows:
\begin{equation}
\varepsilon(E, t, \theta, \alpha, x,y) = \varepsilon_R(E,\theta,\alpha,x,y) \, \varepsilon_a(E,\theta,\alpha) \, \varepsilon_{i}(E,x,y,t) \,.
\end{equation}
The first term, $\varepsilon_{R}$, denotes the effective area of the detector for the energy $E$ and arrival directions ($\theta$,$\alpha$).
Since the Tunka-Rex array is almost radially symmetric, this term can be written in the following form:
\begin{equation}
\varepsilon_R(E,\theta,\alpha,x,y) = \Theta(R_{E,\theta,\alpha}^2 - x^2 - y^2)\,,
\end{equation}
where $R_{E,\theta,\alpha}$ is the effective radius of the array for the energy $E$ and arrival directions ($\theta$,$\alpha$) 
(see Fig.~\ref{fig:eff-radius}).
The effective radius of detector is defined as right boundary of the ``plateau" in efficiency.
For the present work this radius was fixed on the value of 400~m during all calculations.

\begin{figure}[t]
\centering
\includegraphics[width=1.0\linewidth]{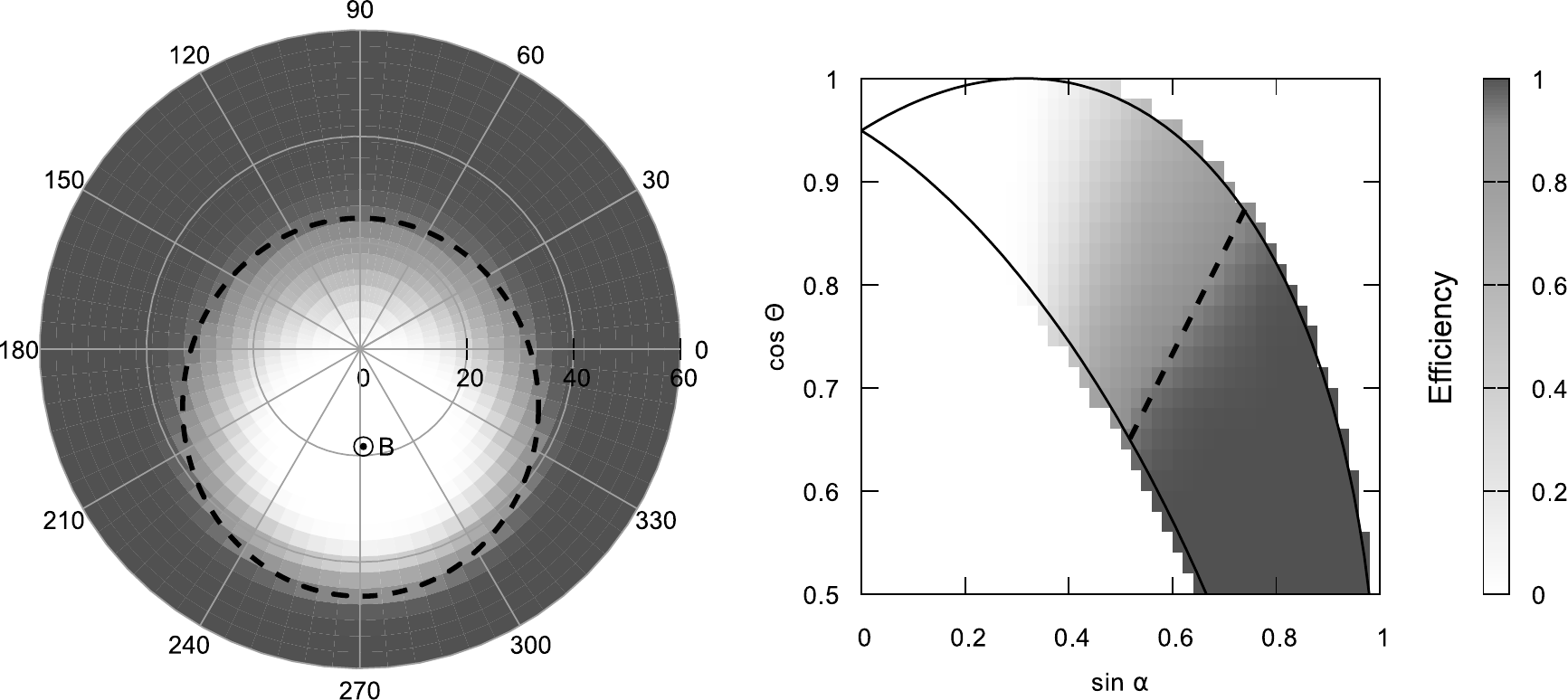}
\caption{Comparison between the total detection efficiency calculated in standard polar coordinates ${(\theta,\phi)}$ and in modified geomagnetic coordinates ${(\cos\theta,\sin\alpha)}$ for an energy of $10^{17.4}$~eV.
The efficiency of 90\% is marked as dashed line between light and dark coloring.
The efficiency cut is more easily described in modified coordinates,
where the solid line indicates the physically allowed range.}
\label{fig:comp_coordsys}
\end{figure}

The second term, $\varepsilon_{a}$ denotes the efficiency depending on the primary energy and arrival direction.
For every energy $E$ we calculate the efficiency as a function of $\cos\theta$ and $\sin\alpha$.
This choice of coordinate system is motivated by strong linear dependences $\varepsilon_{a}(\cos\theta)$ (size of footprint is proportional to $\cos\theta$) 
and $\varepsilon_{a}(\sin\alpha)$ (amplitude of radio emission is proportional to $\sin\alpha$).
By this, the distribution of $\varepsilon_{a}$ in this coordinate system is smooth and can by interpolated without any numerical artifacts.
The comparison between different coordinate systems is given in Fig.~\ref{fig:comp_coordsys}.

The last term, $\varepsilon_i$, denotes the inactive areas due to malfunctions inside the clusters at the time $t$.
These areas have non-trivial shape and can interfere in case of several inactive antenna stations.
We do not consider this term in the present proceeding and plan to parameterize it using Monte Carlo method,
i.e. at the moment the efficiency is overestimated in case of malfunctioning antennas.

\begin{figure}[t]
	\centering
	\includegraphics[width=1.0\linewidth]{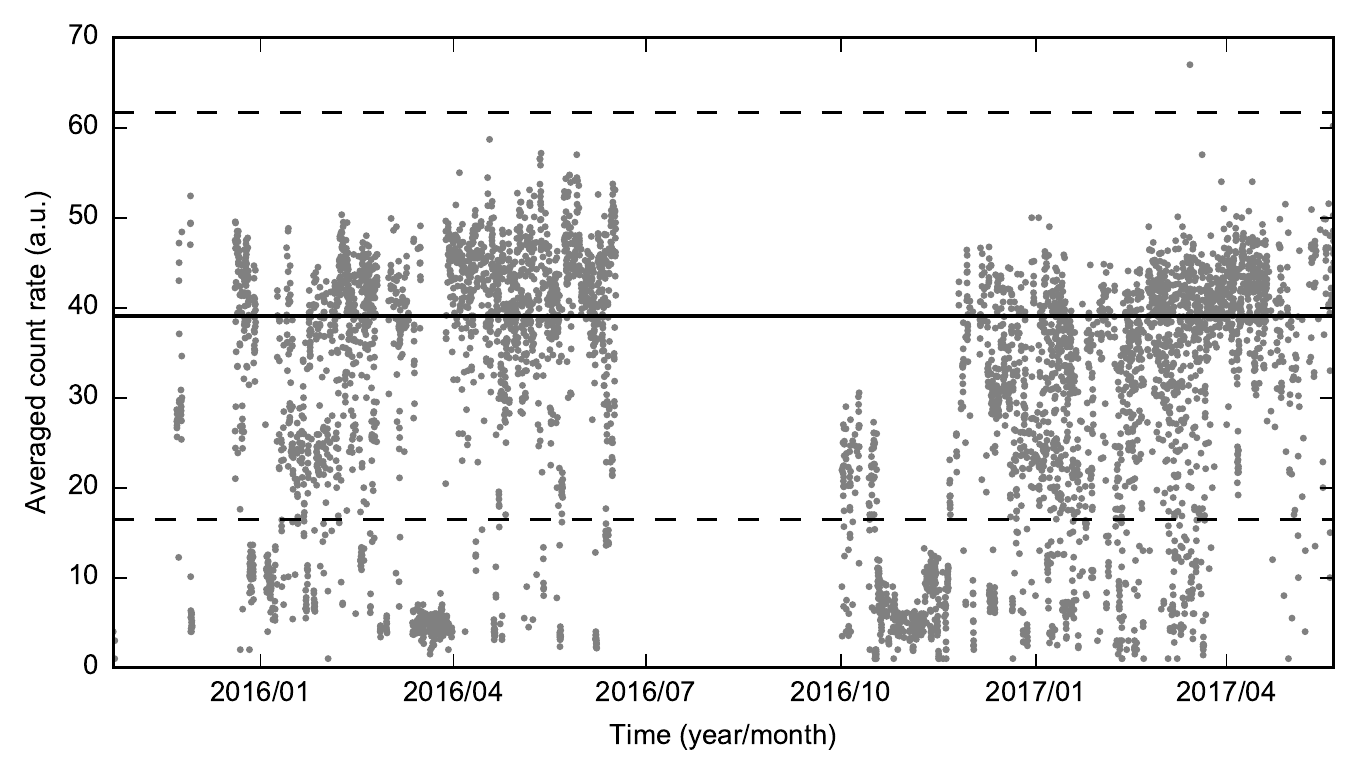}
	\caption{Distribution of the count rates (in arbitrary units) of the whole Tunka-Grande instrument averaged over an hour for two years of observation.
	Only cases where more than three station were triggered were took into account.
	The gap in the middle of 2016 is caused by the summer maintenance. 
	The solid and dashed lines represent the mean value of the count rate $+/-$ two standard deviations, respectively.
	The data out of this band (about 25\,\% of total) are rejected in the analysis.}
	\label{fig:crates}
\end{figure}

\section{The instrument operation time and the uptime of an antenna station}
The Tunka-Rex antenna array receives the triggers for readout from the two TAIGA cosmic-ray instruments: Tunka-133 and Tunka-Grande.
The former instrument is well studied for almost ten years.
The latter instrument has recently been deployed and the knowledge about its operation features and possible problems are fairly modest. 
Thus, we took special care of the Tunka-Rex operation when triggered by Tunka-Grande.

One of the most simple, informative and model-independent parameters
is the count rate.
The cosmic rays flux is approximately constant in our energy range,
which means the count rate of ideal cosmic-ray detector obeys Poisson distribution. 
For real detectors this distribution is modified by processed unrelated to cosmic rays.
Since the instrument have an occasional problems with electronics the main Poisson distribution representing the ``normal'' operation will be accompany with
another distribution on the side of the main one.
So, the parameters of the count rate distribution during the ``normal'' operation can be used for monitoring the instrument state.
Data collected during a period with non-Poisson distributed count rates are excluded from the analysis. 
An example of applying this approach used by us is shown in Fig.~\ref{fig:crates}.

As for the count rates, the effective area of the real instrument varies with time under the influence of diverse factors.
This can be because of hardware malfunctions or saturation due to noise or cross-talks, as well as data corruption during readout and transmission.
The preliminary studies reveals, that about 15\% of detector are affected by this issue.
This does not lead to the exclusion of operation period, but to modification of the aperture.

\section{Comparison with simulations and early data}
\begin{figure}[t]
\centering
\includegraphics[height=0.4\linewidth]{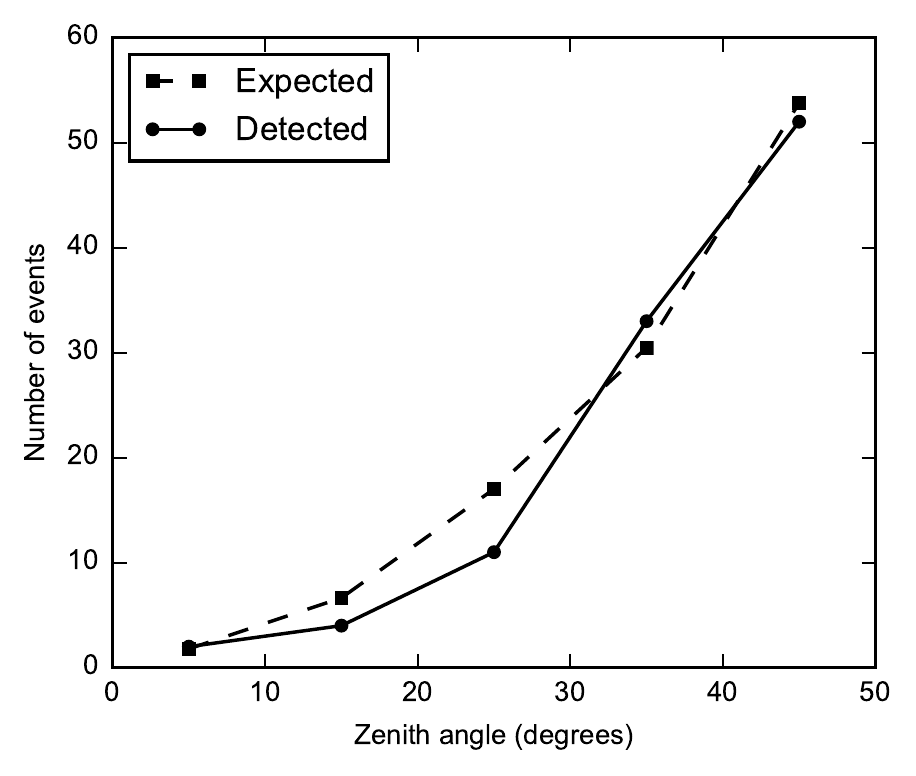}~
\includegraphics[height=0.4\linewidth]{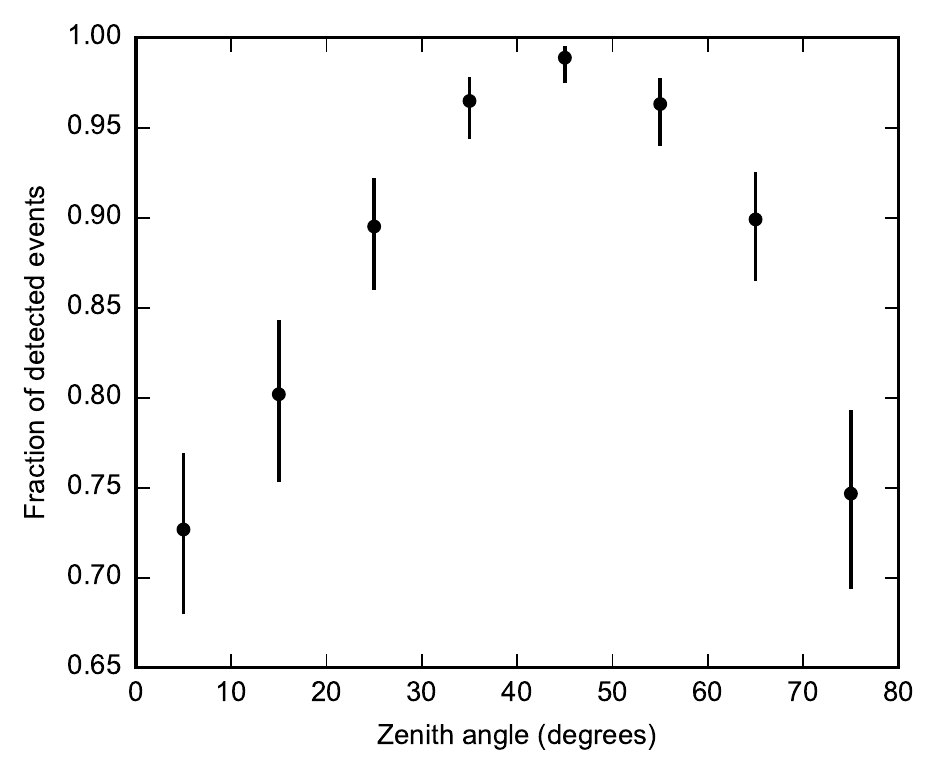}
\caption{Comparison between the detection predicted by the model and given after the Tunka-Rex standard reconstruction.
Left plot shows measurements in 2012-2014, the right plot shows the fraction between detected and expected events for the CoREAS simulations of the layout with 2 antennas per cluster.}
\label{fig:eff_comp}
\end{figure}

To test the performance of our model, we compare its predictions with the data acquired in the years 2012-2014, when the layout of the Tunka-Rex array consisted of 1 antenna station per cluster as well as to CoREAS simulations~\cite{Huege:2013vt} for
the layout with 2 antenna stations per cluster.
To calculate the efficiency, we use the energy and geometry from Tunka-133 for the measurements and the true values from the simulations.
Thus, the expected number of events is:
\begin{equation}
N_{\mathrm{exp}} = \sum\limits_{i = 1}^{N_{\mathrm{tot}}}\varepsilon(E_i, t_i, \theta_i,\phi_i, x_i,y_i)\,,
\end{equation}
i.e. if we use as input $N_{\mathrm{tot}}=200$ events with an average efficiency of $\varepsilon = 0.5$, we expect to detect $N_{\mathrm{exp}} = 100$ events with Tunka-Rex.
The obtained results are shown in Fig~\ref{fig:eff_comp}.
One can see good agreement for the data of 2012-2014, which is expected, since the model was tuned against these measurements.
This tuning was obtained using events with zenith angles $30^\circ$~--~$50^\circ$.
The model overestimates the efficiency for zenith angles less than $30^\circ$ and greater than $60^\circ$.
That is explained by the simplicity of the model, particularly neglecting distance to the shower maximum and antenna response pattern.
Nevertheless, there is no unexpected features in the behavior of our model, and we plan to improve it in near time.

\section{Conclusion}
We have developed a model and formalism for the estimation of the aperture and exposure of modern radio arrays for the detection of cosmic-ray air-showers.
This model reduces the complex radio emission to only one parameter, namely, the size of the radio footprint of the air-shower as function of energy and arrival direction.
Our formalism allows us to estimate the aperture of radio timing array taking into account features of the radio emission from air-showers.

\begin{figure}[t]
\centering
\includegraphics[width=0.49\linewidth]{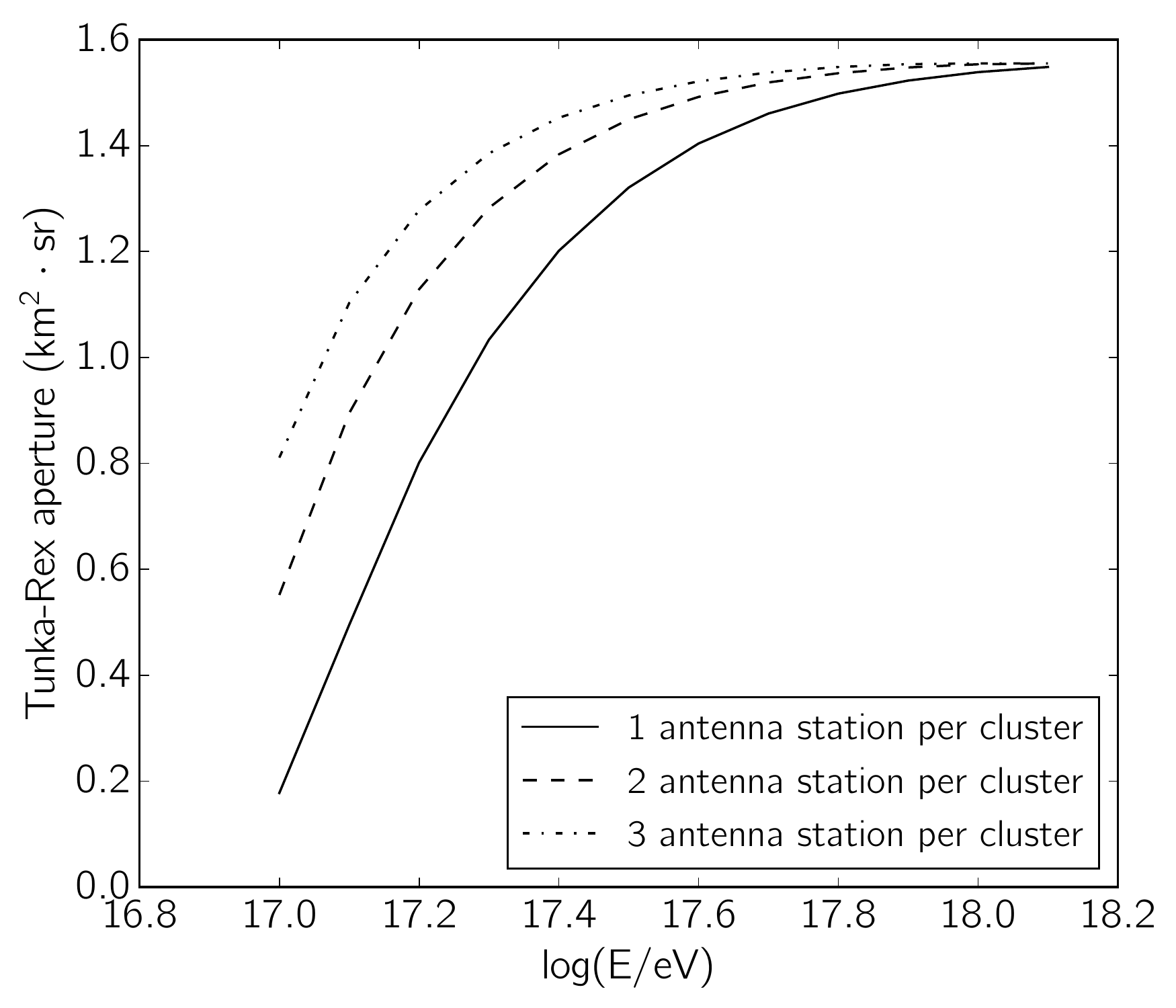}~
\includegraphics[width=0.5\linewidth]{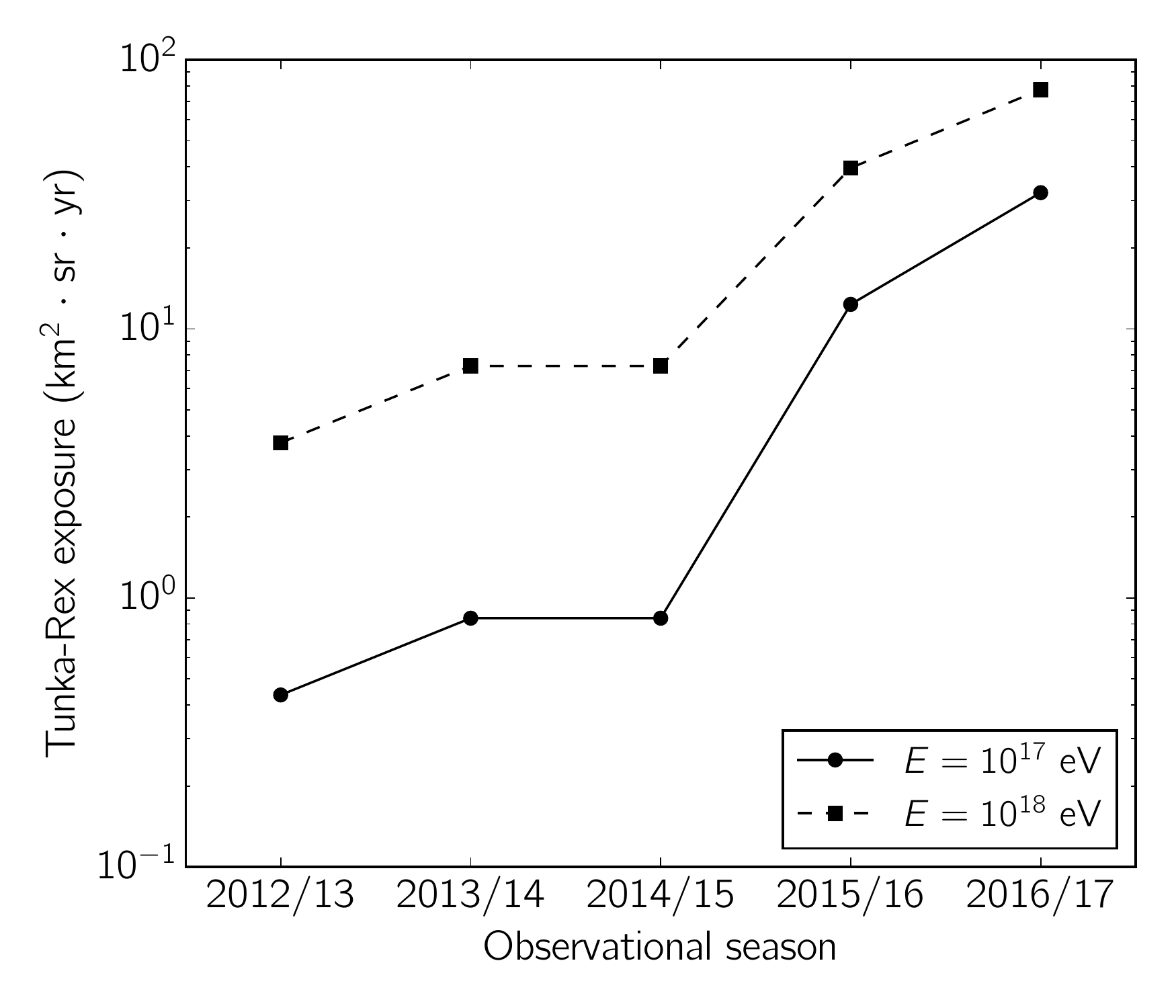}
\caption{Estimated aperture of the Tunka-Rex detector for zenith angles of $0-60^\circ$ in assumption of ideal hardware performance (left) and cumulative exposure for all seasons of observations for present moment (right).
One can see the enhancements after the upgrades of the instrument.}
\label{fig:ap_exp}
\end{figure}

Using this approach the efficiency of the Tunka-Rex detector has been studied.
Cross-check with real data and simulations has shown that the behavior of the Tunka-Rex efficiency can be estimated in a wide range of zenith angles.
By this we estimate the aperture and cumulative exposure of the instrument (see Fig.~\ref{fig:ap_exp}).
However, the current uncertainties are still too large for an accurate measurement of the energy spectrum.
Ideally, the uncertainty of the exposure should be small compared to the energy scale uncertainty of Tunka-Rex of about 20\%.

Consequently, we plan to improve our model and include parameterization taking into account distance to shower maximum and antenna response pattern, as well as correction of aperture for hardware malfunctions.

\section*{Acknowledgements}
The construction of Tunka-Rex was funded by the German Helmholtz association and the Russian Foundation for Basic Research (grant HRJRG-303).
This work has been supported by the Helmholtz Alliance for Astroparticle Physics (HAP),
by Deutsche Forschungsgemeinschaft (DFG grant SCHR 1480/1-1),
by the Russian Federation Ministry of Education and Science (projects 14.B25.31.0010, 2017-14-595-0001, 3.9678.2017/BCh, N3.904.2017/PCh),
by the Russian Foundation for Basic Research (grants 16-02-00738, 16-32-00329, 17-02-00905),
and by grant 15-12-20022 of the Russian Science Foundation (section~3). 

\bibliographystyle{JHEP}
\bibliography{references}

\end{document}